\documentclass[12pt,preprint,apj]{emulateapj}

\usepackage{epstopdf}

\newcommand{\msigma}{M$_{\rm BH}$-$\sigma_*$}
\newcommand{\mbh}{M_{\rm BH}}

\newcommand{\sigstar}{\sigma_{*}}
\newcommand{\sigcorr}{\sigma_{R}}
\newcommand{\siguncorr}{\sigma_{R}(uncor)}
\newcommand{\sigopt}{\sigma_{opt}}
\newcommand{\sigir}{\sigma_{IR}}
\newcommand{\reff}{R_{\rm e}}

\shorttitle{}
\shortauthors{Kang et al.}

\begin{document}

\title{CALIBRATING STELLAR VELOCITY DISPERSIONS BASED ON SPATIALLY-RESOLVED $H$-BAND SPECTRA
FOR IMPROVING THE \msigma\ RELATION}

\author{WOL-RANG KANG$^{1}$}
\author{JONG-HAK WOO$^{1}$\altaffilmark{,7}}
\author{ANDREAS SCHULZE$^{2}$}
\author{DOMINIK A. RIECHERS$^{3,4}$}
\author{SANG CHUL KIM$^{5}$}
\author{DAESEONG PARK$^{1}$}
\author{VERNESA SMOLCIC$^{6}$}

\affil{$^{1}$Astronomy Program, Department of Physics and Astronomy, Seoul National University, 1 Gwanak-ro Gwanak-gu, Seoul, 151-742, Republic of Korea; woo@astro.snu.ac.kr}
\affil{$^{2}$Kavli Institute for Astronomy and Astrophysics, Peking University, 100871 Beijing, China}
\affil{$^{3}$Astronomy Department, Cornell University, 220 Space Science Building, Ithaca, NY 14853, USA}
\affil{$^{4}$Astronomy Department, California Institute of Technology, MC 249-17, 1200 East California Boulevard, Pasadena, CA 91125, USA}
\affil{$^{5}$Korea Astronomy and Space Science Institute, Daejeon 305-348, Republic of Korea}
\affil{$^{6}$Physics Department, University of Zagreb, Bijeni\v{c}ka cesta 32, 10002 Zagreb, Croatia}
\altaffiltext{7}{Author to whom any correspondence should be addressed}

\begin{abstract}

To calibrate stellar velocity dispersion measurements from optical and near-IR stellar lines,
and to improve the black hole mass (M$_{\rm BH}$)- stellar velocity dispersion ($\sigma_*$)
relation, we measure $\sigstar$ based on high quality $H$-band spectra 
for a sample of 31 nearby galaxies,
for which dynamical $\mbh$ is available in the literature.
By comparing velocity dispersions measured from stellar lines in the $H$-band
with those measured from optical stellar lines,
we find no significant difference, suggesting that optical and near-IR 
stellar lines represent the same kinematics and that dust effect is negligible
for early-type galaxies.
Based on the spatially-resolved rotation and velocity dispersion measurements along the major axis
of each galaxy, we find that a rotating stellar disk is present for 80\% of galaxies 
in the sample. For galaxies with a rotation component, $\sigma_*$ measured from a single 
aperture spectrum can vary by up to $\sim$20\%, depending on the size of the adopted extraction aperture.
To correct for the rotational broadening, we derive luminosity-weighted $\sigstar$
within the effective radius of each galaxy, providing uniformly measured
velocity dispersions to improve the \msigma\ relation.

\end{abstract}


%==============================================================================================================

\section{Introduction}

The black hole mass ($\mbh$) correlation with host galaxy properties
has been one of the main issues in understanding galaxy evolution and
black hole growth. In particular, the relatively tight correlation between $\mbh$
and stellar velocity dispersion (\msigma) has been reported for nearby galaxies 
with dynamically measured  $\mbh$ \citep{FM+00,geb+00}, as well as 
present-day active galaxies with $\mbh$ determined from reverberation-mapping results
\citep{onken+04,woo+10}.
While early studies claimed a remarkably tight \msigma\ relation with its intrinsic 
scatter below $\sim$0.3 dex \citep[e.g.,][]{T+02}, recent studies
showed a larger intrinsic scatter and a steeper slope due to the increased sample size,
inclusion of more diverse galaxies, i.e., late-type and pseudo bulge galaxies,
and the improvements of $\mbh$ measurements based on better dynamical modeling 
and data \citep{FF+05,gra+08,gul+09a,McC+11,McC+12}.

In understanding BH-galaxy coevolution, 
the present-day \msigma\ relation sets a local calibration point 
as most observational studies investigated cosmic evolution of the \msigma\ relation 
by measuring an offset from the local relationship 
\citep[e.g.,][]{woo+06,woo+08,bennert+10,bennert+11b}.
At the same time, the present-day \msigma\ relation has been used 
for calibrating the $\mbh$ of active galactic nuclei (AGN), which is determined from the kinematics of 
sub-pc scale broad-emission line region. 
The unknown viral factor for converting
the line-of-sight velocity of broad-line region gas to the intrinsic velocity,
has been empirically determined by matching the \msigma\ relation of quiescent 
and active galaxies at z $\sim$0 \citep{onken+04,woo+10,park+12}.
Thus, defining the \msigma\ relation in the local universe is of importance to unveil
the nature of BH-galaxy coevolution. 

Stellar kinematics studies based on the near-IR stellar lines
became powerful as near-IR spectrographs combined with laser-guide star adaptive optics 
provides the best spatial resolution for the ground-based facilities
\citep[e.g.,][]{watson+08}. 
Moreover, measuring $\sigstar$ in the near-IR is more promising for AGN host galaxies 
since AGN-to-star flux ratios are much more favorable in the near-IR 
\citep{dasyra+07,watson+08, woo+10}
while it is almost impossible to measure $\sigstar$ in the optical for host galaxies 
of high luminosity QSOs.

Despite the increasing usage of near-IR spectra for probing stellar kinematics, 
a proper comparison between optical and near-IR measurements is still lacking.  
By measuring velocity dispersion of 25 early-type galaxies based on the CO absorption band head 
at 2.29\micron\ in the $K$-band, \citet{SG+03} claimed that 
velocity dispersion measured from near-IR stellar lines was systematically smaller 
by 10-30$\%$ than that measured from optical stellar lines.
In contrast, \citet{r&f10} reported that optical and near-IR velocity dispersions
were consistent for a sample of 23 early-type galaxies, 
by comparing $\sigstar$ measured from the CO band heads in the $K$-band, with 
$\sigstar$ measured from the CaII triplet.
\citet{vdb+11} also measured velocity dispersion based on the CO band heads
for a sample of 22 early-type galaxies, and presented consistent results with
respect to
optical $\sigstar$. The discrepancy among various studies may have resulted from
the systematic uncertainties of the velocity dispersion measurements 
since the line dispersion was measured from intrinsically broad CO band heads in the $K$-band 
and template mismatch could be very strong \citep[see][]{SG+03}.
In contrast, the $H$-band spectral range ($\sim$1.6-$\sim$1.7\micron) contains many 
more stellar lines, e.g., Si I, CO, and Mg I than the $K$-band,
and is possibly less susceptible to template mismatch although the presence 
of strong sky OH lines is a downside. To utilize the $H$-band stellar lines for studying
stellar kinematics, a proper comparison is required between $\sigstar$ measured from $H$-band spectra
with that measured from optical spectra stellar lines.

To derive reliable $\sigstar$ to represent the kinematics 
of the pressure-supported bulge or spheroidal component, the effect of the rotation component 
should be corrected for. In the case of galaxies with a rotating stellar disk, 
the line-of-sight velocity dispersion can be easily overestimated due to rotational broadening
if a large aperture is used to extract a spectrum \citep[e.g.,][]{bennert+11a,harris+12}.
The effect of rotational broadening is stronger for more edge-on stellar disks,
potentially producing systematic bias.  
Thus, it is important to correct for rotational broadening. 

In this paper, we measure the stellar velocity dispersion of 31 nearby galaxies
using high quality $H$-band spectra.
We compare $\sigstar$ measurements based on stellar lines in the $H$-band with optical $\sigstar$ 
measurements from the literature. We also correct for the rotation and aperture effect based on the spatially 
resolved kinematics measurements to improve the \msigma\
relation. 
The paper is organized as follows.
We describe sample selection, observations and data reduction in \S~\ref{section:obs}. 
In \S~3, we present $\sigstar$ measurements, and the effects of
rotation and aperture size. 
In \S~4, we compare our $H$-band $\sigstar$ measurements
with optical $\sigstar$ from the literature, and derive the \msigma\ relation
for early-type galaxies based on the rotation-corrected $\sigstar$.
The main results are summarized in \S~\ref{section:sum}.

%==============================================================================================================

\section{Observations and Data Reduction} \label{section:obs}

\subsection{Sample Selection and Observations}

To directly compare optical and near-IR stellar velocity dispersions and to calibrate
the \msigma\ relation, we select 31 nearby galaxies from the \msigma\ sample
\citep[e.g.,][]{gul+09a}, for which dynamical $\mbh$ measurements and 
optical $\sigstar$ measurements are available.
The sample is mainly composed of early-type galaxies (20 ellipticals, 8 lenticulars 
and 3 spirals) and spans a wide range in $\sigstar$ 
from 67 km s$^{-1}$ to 385 km s$^{-1}$ as listed in Table~\ref{table1}.
Also, it covers three orders of magnitude in $\mbh$ and 
constitutes about half the sample size of galaxies with dynamical $\mbh$ 
measurements \citep{McC+13}.

Observations were performed at the Palomar Hale 5~m telescope using the near-IR 
spectrograph TripleSpec, simultaneously covering the wavelength range from 1.0 $\micron$ to 2.4 $\micron$. In this work we only employ the $H$-band spectra centered at $\sim$1.7\micron, as it covers many stellar absorption lines suitable for the $\sigstar$ measurement. We place an 1$\arcsec$ $\times$ 30$\arcsec$ long-slit along the major axis of each galaxy.
The spectral resolution of TripleSpec is R = 2500 - 2700, corresponding to a Gaussian dispersion $\sim$50 km s$^{-1}$. As the lowest optical $\sigstar$ is 67 km s$^{-1}$ (for NGC 7457), this spectral resolution is suitable for our study.

For sky subtraction, in particular for the strong OH sky emission lines, 
we also observed blank sky, offset by several arc minutes from each galaxy, 
since the size of each galaxy is larger than the slit length and fills the 
entire slit. We divide the total exposure time into segments of 200 second 
exposures to avoid saturation in the $K$-band. The total on-source exposure time
ranges from 600 to 1000 seconds depending on the magnitude of individual galaxies (see Table 1).
We observed several A0V stars each night to correct for telluric lines.
We also observed 11 K- and M-type giant stars as velocity templates for the stellar velocity 
dispersion determination. 

\subsection{Data Reduction}

We performed standard data reduction, i.e., bias subtraction, flat-fielding and wavelength calibration using a series of IRAF scripts, then extracted one-dimensional 
spectra using various extraction windows.
For telluric absorption correction, we constructed a telluric line template for 
each observing night, based on the spectra of A0V stars observed during the night.  For each A0V star, we fitted their Brackett lines with double Gaussians and normalized the spectra by its continuum. Dividing the observed A0V star spectrum by this model spectrum provides a telluric template. We combined all telluric 
templates to construct a mean template for a given night. Then, we used the template
to correct the galaxy spectra for telluric absorption lines. 
%Using a mean template for each night removes small variations between individual stars.

To investigate the effect of galaxy rotation on stellar velocity dispersion 
measurements, we extracted spatially resolved spectra from a number 
of small extraction windows (4-16 pixels) along the galaxy's major axis, 
which were allowed to overlap with each other and to slightly increase at larger radii 
for obtaining better signal-to-noise. 
The resolution for spatial binning depends on the distance to each galaxy.
Typically $\sim$10 spectra were extracted within a fraction of the $\reff$ (see Section 3.5 for details).
 
A series of single-aperture spectra were also extracted using various aperture sizes, 
in order to investigate the aperture effect.
Many previous studies used spatially unresolved $\sigstar$ measurements, 
which were affected by line broadening due to galaxy rotation. This leads to an overestimation of the galaxy's $\sigstar$, depending on how much rotation is included in
the extraction aperture. Thus, the choice of different aperture sizes can affect  
the $\sigstar$ measurement as presented in \S~3.4.
In contrast, we can correct the $\sigstar$ measurement for the 
rotational broadening using our spatially resolved spectra. 
Details on this correction are presented in \S~3.4 and \S~3.5.

%%% Table 1%%%%%%%%%%%%%%%%%%%%%%%%%%%%%%%%%%%%%%%%%%%%%%%%%
\begin{deluxetable*}{cccccccccccccc}[ht]
%\rotate
\tablecolumns{14}
\tablewidth{0pc}
\tablecaption{Sample Selection and Observing Log}
\tablehead{
\colhead{Galaxy} & \colhead{RA} & \colhead{DEC} & \colhead{Type} &  \colhead{Dist.} & \colhead{Spatial} & \colhead{$R_e$} & \colhead{Ref.} & \colhead{$M_{\rm BH}$} & \colhead{Ref.}  & \colhead{UT Date} & \colhead{$T_{EXP}$} & \colhead{S/N} & \colhead{PA}\\
\colhead{} & \colhead{(J2000)} & \colhead{(J2000)} & \colhead{} &  \colhead{} & \colhead{Scale} & \colhead{} & \colhead{} & \colhead{} & \colhead{}  & \colhead{} & \colhead{} & \colhead{} & \colhead{}\\
\colhead{ }&\colhead{ }&\colhead{ }&\colhead{ }& \colhead{(Mpc)} & \colhead{(kpc/1{\arcsec})} & \colhead{(kpc)} & \colhead{ } & \colhead{($10^8\, M_{\sun}$)} &\colhead{ }&\colhead{ }&\colhead{(s)}&\colhead{ }& \colhead{{(\degr)}}\\
\colhead{(1)}&\colhead{(2)}&\colhead{(3)}&\colhead{(4)}&\colhead{(5)}&\colhead{(6)}&\colhead{(7)}&\colhead{(8)}&\colhead{(9)}&\colhead{(10)}&\colhead{(11)}&\colhead{(12)}&\colhead{(13)}&\colhead{(14)}
}
\startdata
N221   & 00 42 41.87 & $+$40 51 57.2 &  E2        & 0.86   & 0.0037 & 0.24 &1  & $0.026{\pm0.005}$           & 25, 6     &  10 Jan 1   &600   &414   & 170\\
N821   & 02 08 21.04 & $+$10 59 41.1 &  E4        & 25.5   & 0.1202 & 15.7 &2  & $1.7{\pm0.7}$                     & 25, 7     &  10 Jan 1   &600   &133   & 25\\
N1023 & 02 40 23.90 & $+$39 03 46.3 &  SB0     & 12.1   & 0.0442 & 1.1   &2  & $0.4{\pm0.04}$                   & 25, 8     &  10 Jan 1   &600   &320   & 87\\
N1068 & 02 42 40.83 & $-$00 00 48.4  &  Sb        & 15.4   & 0.0788 & 2.9   &2  & $0.086{\pm0.003}$           & 5, 9     &  10 Jan 1   &600   &455   & 13\\
N2778 & 09 12 24.35 & $+$35 01 39.4 &  E2        & 24.2   & 0.1420 & 3.0   &1  & $0.16_{-0.102}^{0.09}$    & 4         &  09 May 22   &600   &158   & 40\\
N2787 & 09 19 18.90 & $+$69 12 11.9 &  SB0     & 7.9      & 0.0482 & 0.27 &2  & $0.41_{-0.05}^{0.04}$      & 25, 10  &  09 May 25   &600   &247   & 109\\
N3031 & 09 55 33.17 & $+$69 03 55.1 &  Sb        & 4.1      & 0.0179 & 3.0   &2  & $0.8_{-0.11}^{0.2}$          & 25, 11  &  10 Jan 1   &600   &393   & 149\\
N3115 & 10 05 13.80 & $-$07 43 08.0  &  S0        & 10.2   & 0.0460 & 2.9   &2  & $8.9_{-2.7}^{+5.1}$           & 25, 12  &  10 Jan 1   &600   &362   & 40\\
N3245 & 10 27 18.52 & $+$28 30 24.8 &  S0        & 22.1   & 0.0911 & 1.0   &2  & $2.1_{-0.6}^{+0.5}$            & 25, 13  &  09 May 22   &600   &247   & 177\\
N3377 & 10 47 42.36 & $+$13 59 08.8 &  E6        &11.7    & 0.0461 & 3.9   &2  & $1.8{\pm0.9}$                     & 25, 7    &  10 Jan 1   &600   &244   & 41\\
N3379 & 10 47 49.75 & $+$12 34 54.6 &  E0        & 11.7   & 0.0631 & 2.7   &2  & $4.2_{-1.1}^{+1.0}$           & 25, 14  &  09 May 22   &600   &307   & 73\\
N3384 & 10 48 16.90 & $+$12 37 42.9 &  SB0      & 11.7   & 0.0488 & 0.48 &2  & $0.11{\pm0.05}$                & 25, 7    &  10 Jan 1   &600   &254   & 53\\
N3607 & 16 12 54.64 & $+$18 03 06.3 &  E1        & 19.9   & 0.0665 & 4.3   &3  & $1.4_{-0.5}^{+0.4}$            & 25, 12  & 10 Jan 1   &800   &229   & 125\\
N3608 & 11 16 59.07 & $+$18 08 54.6 &  E1        & 23.0   & 0.0869 & 3.9   &2  & $4.7{\pm1.0}$                     & 25, 7    &  09 May 25   &800   &252   & 75\\
N4258 & 12 18 57.54 & $+$47 18 14.3 &  SABbc & 7.2     & 0.0310 & 0.66 &2  & $0.367{\pm0.001}$           & 25, 16  &  09 Mar 5   &600   &229   & 150\\
N4261 & 12 19 23.21 & $+$05 49 29.7 &  E2        & 33.4   & 0.1551 & 5.8   &2  & $5.3{\pm1.1}$                     & 25, 17  &  09 May 25   &600   &203   & 160\\
N4291 & 12 20 17.60 & $+$75 22 15.0 &  E2        & 25.0   & 0.1218 & 2.0   &2  & $9.8{\pm3.1}$                     & 25, 7    &  10 Jan 1   &800   &164   & 110\\
N4342 & 12 23 39.12 & $+$07 03 12.9 &  S0        & 18.0   & 0.0520 & 0.21 &2  & $4.6_{-1.5}^{+2.6}$           & 25, 18  &  09 Mar 5   &800   &104   & 168\\
N4374 & 12 25 03.74 & $+$12 53 13.1 &  E1        & 17.0   & 0.0735 & 7.8   &2  & $9.2_{-0.8}^{+1.0}$           & 25, 19  &  10 Jan 1   &600   &250   & 122.5\\
N4459 & 12 29 00.13 & $+$13 58 42.5 &  E2        & 17.0   & 0.0839 & 13.7 &2  & $0.7_{-0.14}^{+0.13}$      & 25, 10  &  09 Mar 5   &600   &206   & 110\\
N4473 & 12 29 48.95 & $+$13 25 46.1 &  E4        & 17.0   & 0.1555 & 2.1   &2  & $0.89_{-0.44}^{+0.45}$    & 25, 7    &  09 Mar 5   &600   &170   & 100\\
N4486 & 12 30 49.42 & $+$12 23 28.0 &  E1        & 17.0   & 0.0906 & 6.0   &2  & $62.0_{-4.0}^{+3.0}$         & 25, 20  &  09 May 22   &600   &168   & 153\\
N4564 & 12 36 27.01 & $+$11 26 18.8 &  S0        & 17.0   & 0.0791 & 3.0   &1  & $0.88{\pm0.24}$                 & 25, 7    &  09 May 25   &600   &272   & 47\\
N4596 & 12 39 56.16 & $+$10 10 32.4 &  SB0     & 18.0   & 0.1296 & 1.5   &2  & $0.84_{-0.25}^{+0.36}$     & 25, 10  &  09 May 25   &600   &208   & 75\\
N4649 & 12 43 40.19 & $+$11 33 08.9 &  E2        & 16.5   & 0.0774 & 7.2   &2  & $47.0_{-10.0}^{+11.0}$     & 25, 21  &  09 May 22   &600   &221   & 105\\
N4697 & 12 48 35.70 & $-$05 48 03.0  &  E6        & 12.4   & 0.0860 & 6.9   &2  & $2.0{\pm0.2}$                      & 25, 7    &  10 Jan 1   &600   &260   & 70\\
N4742 & 12 51 47.92 & $-$10 27 17.1  &  E4        & 16.4   & 0.0880 & 1.6   &2  & $0.14{\pm0.05}$                 & 24      &  09 May 25   &600   &355   & 75\\
N5845 & 15 06 00.90 & $+$01 38 01.4 &  E3        & 28.7   & 0.1005 & 0.42 &2  & $4.9_{-1.6}^{+1.5}$            & 25, 7     & 09 May 22   &600   &274   & 141\\
N6251 & 16 32 31.97 & $+$82 32 16.4 &  E1        & 106.0 & 0.5134 & 10.0 &2  & $6.0{\pm2.0}$                     & 25, 22  & 09 May 22   &600   &141   & 21\\
N7052 & 21 18 33.13 & $+$26 26 48.7 &  E3        & 70.9   & 0.3238 & 9.1   &2  & $4.0_{-1.6}^{+2.8}$            & 25, 23  & 09 May 25   &1000 &185   & 62\\
N7457 & 23 01 00.05 & $+$30 08 43.4 &  S0        & 14.0   & 0.0563 & 4.8   &1  & $0.10{\pm0.06}$                 & 5, 7    & 09 May 25   &600   &131   & 125\\
\enddata
\label{table1}
\tablecomments{Col. (1): NGC galaxy catalogue name. Col. (2) Right Ascension. Col. (3): Declination. Col. (4): morphological types. Col. (5): distance. Col. (6): spatial scale. Col. (7)-(8): effective radius and reference. Col. (9)-(10): black hole mass and their reference. Col. (11): observation date. Col. (12): total exposure time. Col. (13): average signal-to-noise ratio within ${\pm 5}$ pixel aperture. Col. (14): Position Angle.\\
References. --- (1) \citet{MH+03}; (2) \citet{gra+08}; (3) \citet{sani+11};  (4) \citet{gul+09a}; (5) \citet{McC+11}; (6) \citet{ver+02} (7) \citet{SG+11}; (8) \citet{bower+01}; (9) \citet{LB+03}; (10) \citet{sarzi+01}; (11) \citet{dev+03}; (12) \citet{EDB99}; (13) \citet{barth+01}; (14) \citet{vdBdZ+10}; (15) \citet{gul+09b}; (16) \citet{her+05}; (17) \citet{FFJ96}; (18) \citet{CvdB99}; (19) \citet{WBS+10}; (20) \citet{geb+11}; (21) \citet{SG+10}; (22) \citet{FF99}; (23) \citet{VV98}; (24) \citet{T+02}; (25) \citet{McC+13}
}
\end{deluxetable*}
%%%%%%%%%%%%%%%%%%%%%%%%%%%%%%%%%%%%%%%%%%%%%%%%%%%%%%%%%%%%

%%% Figure 1%%%%%%%%%%%%%%%%%%%%%%%%%%%%%%%%%%%%%%%%%
\begin{figure}[ht]
\centering
\includegraphics[width = 0.45 \textwidth]{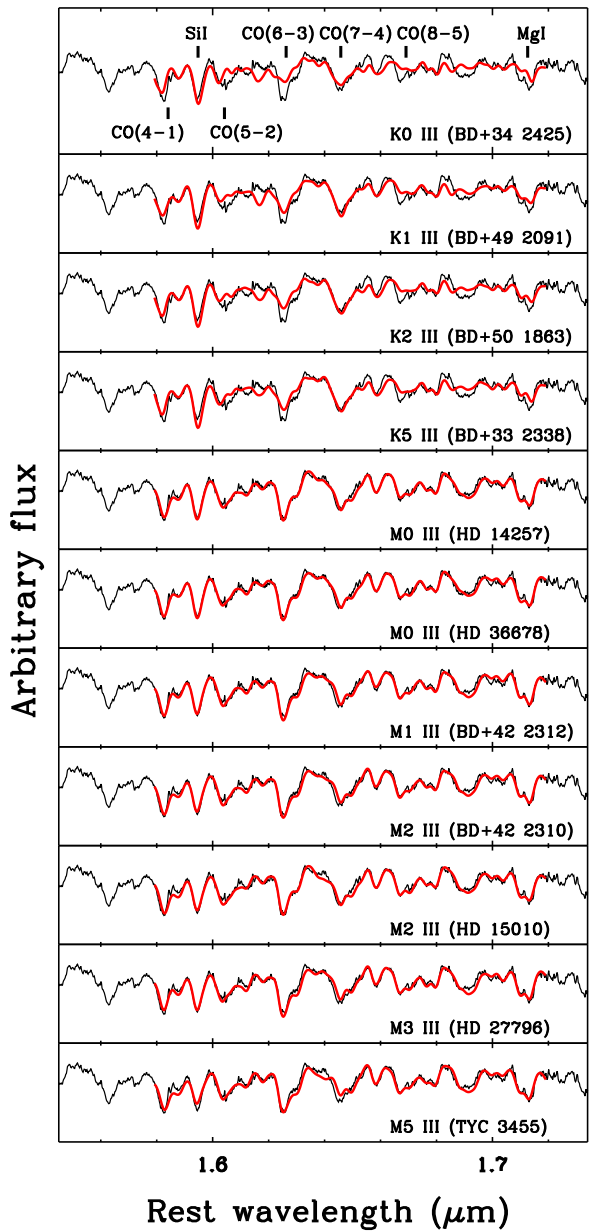}
\caption{Normalized spectra of template stars. The template star spectra (thick red solid line) are compared with the spectra of NGC 1023 (thin black solid line). The template star is broadened with a Gaussian velocity kernel. Individual template stars show different line strengths, particularly for the CO absorption line. The K type star templates provide a poor fit to the galaxy spectrum. In the top panel, we marked several individual stellar lines with black tick masks (from left, CO(4-1), SiI, CO(5-2), CO(6-3), CO(7-4), CO(8-5) and MgI).}
\label{fig1}
\end{figure}
%%%%%%%%%%%%%%%%%%%%%%%%%%%%%%%%%%%%%%%%%%%%%%%%%

%%% Figure 2%%%%%%%%%%%%%%%%%%%%%%%%%%%%%%%%%%%%%%%%%
\begin{figure}
\centering
\includegraphics[width = 0.47 \textwidth]{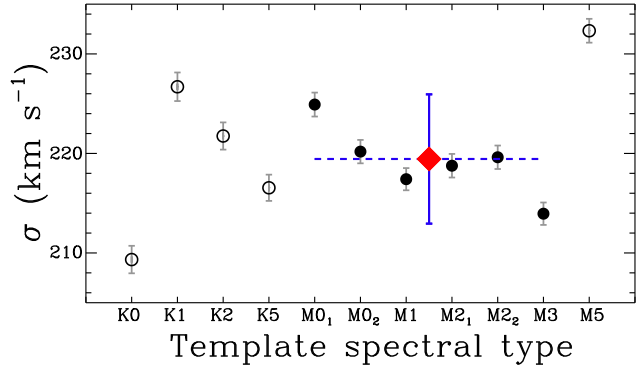}
\caption{Comparison of the measured stellar velocity dispersion of NGC 1023 using different 
template stars. Filled circles denote M0 III, M1 III, M2 III and M3 III type stars, which were used for calculating the mean stellar velocity dispersion. Open circles denote M5 III and K type template stars. The mean $\sigstar$ is given by the red diamond and the standard deviation of the measurements is included in the uncertainty denoted by the blue solid error bar. The blue dashed line indicates the range of  template stars used for calculating the mean stellar velocity dispersion. 
}
\label{fig2}
\end{figure}
%%%%%%%%%%%%%%%%%%%%%%%%%%%%%%%%%%%%%%%%%%%%%%%%%

%%% Figure 3%%%%%%%%%%%%%%%%%%%%%%%%%%%%%%%%%%%%%%%%%
\begin{figure}[ht]
\centering
\includegraphics[width = 0.50 \textwidth]{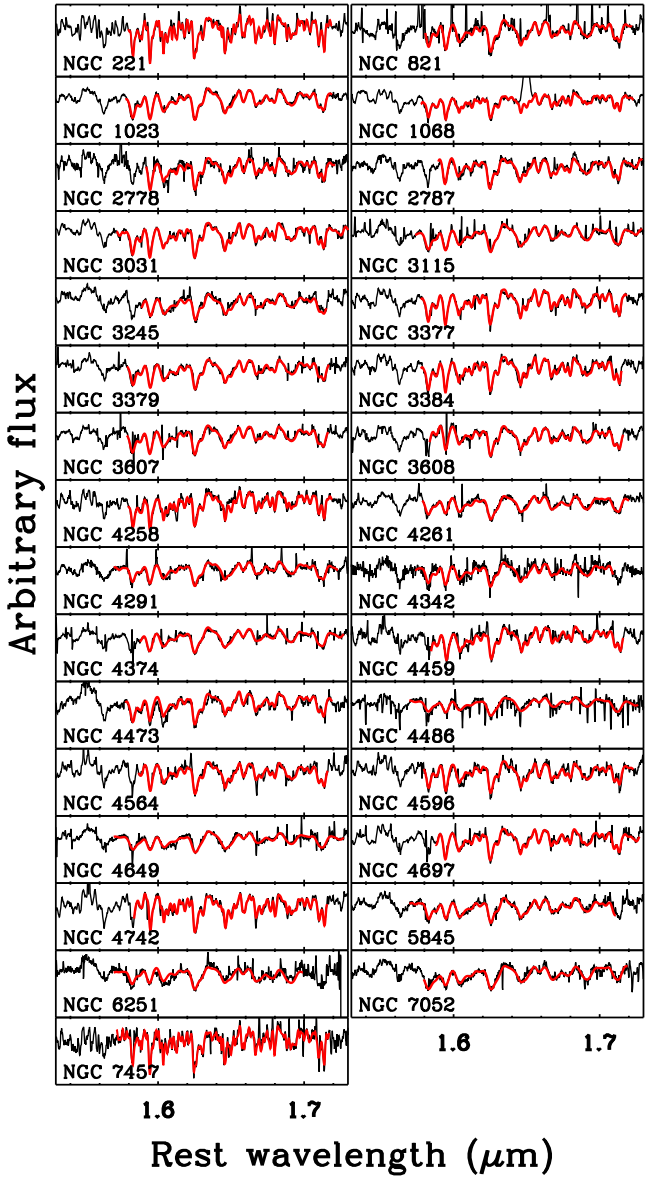}
\caption{Normalized spectra of the 31 galaxies and their best-fit models. 
The  broadened template star spectra (thick red solid line) fit the observed galaxy spectra (thin black solid line) reasonably well. Residuals of OH sky emission lines (sharp feautures in Fig. 1) and the AGN Fe II emission line (e.g., in NGC 1068) were masked out before fitting.}
\label{fig3}
\end{figure}
%%%%%%%%%%%%%%%%%%%%%%%%%%%%%%%%%%%%%%%%%%%%%%%%%%%%%%%%%%%%%

%==============================================================================================================

\section{ANALYSIS} \label{section:anal}

\subsection{Stellar Velocity Dispersion Measurements}
We measured the stellar velocity dispersion of 31 galaxies in the sample using the 
stellar lines in the 1.57 - 1.72 \micron\ range, i.e, CO(4-1) 1.58\micron\, Si~I 1.59\micron\, 
CO(5-2) 1.6\micron\, CO(6-3) 1.62\micron\, CO(7-4) 1.64\micron\, CO(8-5) 1.66\micron, 
and Mg~I 1.71\micron\ (see Figure 1). 
Using the Gauss-Hermite Pixel Fitting software \citep{vdm94,woo04,woo05,woo+06}, 
we performed $\chi^2$ minimization in fitting the galaxy spectra directly in pixel space 
to stellar template spectra broadened by a Gaussian kernel with velocity widths ranging from 
50 to 350 km~s$^{-1}$.
The continua of the spectra of the template stars are fitted with low-order (2-3) polynomials
while the FeII emission line at 1.65$\micron$, bad pixels and residuals from sky line subtraction 
were masked out before the fitting.

\subsection{Template Mismatch}  \label{section:temis}

Since the $\sigstar$ measurement is affected by the choice of template star, it is necessary 
to quantify the uncertainty due to the template mismatch. Using 11 velocity template stars 
of various spectral types, namely, K0 III, K1 III, K2 III, K5III,  two M0 III, M1 III, two M2 III, M3 III, 
and M5 III, which were observed with the same instrumental setup during our observing runs,
we measured and compared $\sigstar$ for individual galaxies in the sample, 
in order to investigate the variation in the $\sigstar$ measurement caused by template mismatch.
Then, we accounted for template mismatch in the determination of  $\sigstar$ by averaging
$\sigstar$ measurements using various template stars.

To compare the overall spectral shapes, we present the spectra of the individual template 
stars, after broadening them with a Gaussian velocity (red thick lines) in Figure \ref{fig1}. 
The observed spectrum of NGC 1023 is overplotted (black lines) to demonstrate
the template mismatch. In the stellar spectra, the line strength of the CO absorption lines 
increases toward later-type stars (from upper panels to lower panels). This trend is in 
particular clearly shown for the CO(6-3) line at 1.62 $\micron$ and also for the MgI line 
at 1.71 $\micron$. On the other hand, the SiI line strength shows no strong variation
with spectral type. The comparison in Fig. 2 clearly shows that spectra of K type stars 
provide a poor match to the observed galaxy spectrum in this wavelength range, while spectra
of M type stars can fit the observed galaxy spectrum reasonably well.

In Figure \ref{fig2}, we compare the multiple measurements of $\sigstar$  of NGC 1023,
using each template star for the fitting.
As expected from Figure 1, the $\sigstar$ measured from M-type stars shows small variation, 
while $\sigstar$ measured from K-type stars exhibits a larger scatter, suggesting that M-type 
stars provide a fair representation of the luminosity-weighted stellar population in the 
$H$-band. Therefore, we excluded the measurements from K-type stars and calculated the mean 
$\sigstar$ based on the 6 M-type stars. Since the M5 III star shows slightly different
line shapes compared to other M-type stars (see Figure 1), 
we also excluded the measurement based on the M5 III template. 
After calculating the standard deviation of the measurements from 6 M-type stars
as the uncertainty of template mismatch, we added the uncertainty of template
mismatch to the mean measurement errors from 6 M-type stars in quadrature,
in order to determine the uncertainty of $\sigstar$.
For example, the red diamond in Figure \ref{fig2} indicates the mean $\sigstar$, derived 
from 6 template stars and its uncertainty.
In Figure \ref{fig3}, we present the normalized observed spectrum (black solid line)
of each galaxy in the sample, overplotted with the best-fit model (red solid line).

\subsection{Spatially Resolved Stellar Velocity Dispersions} \label{section:sig}

\begin{figure*}[htp]
\centering
\includegraphics[width = \textwidth]{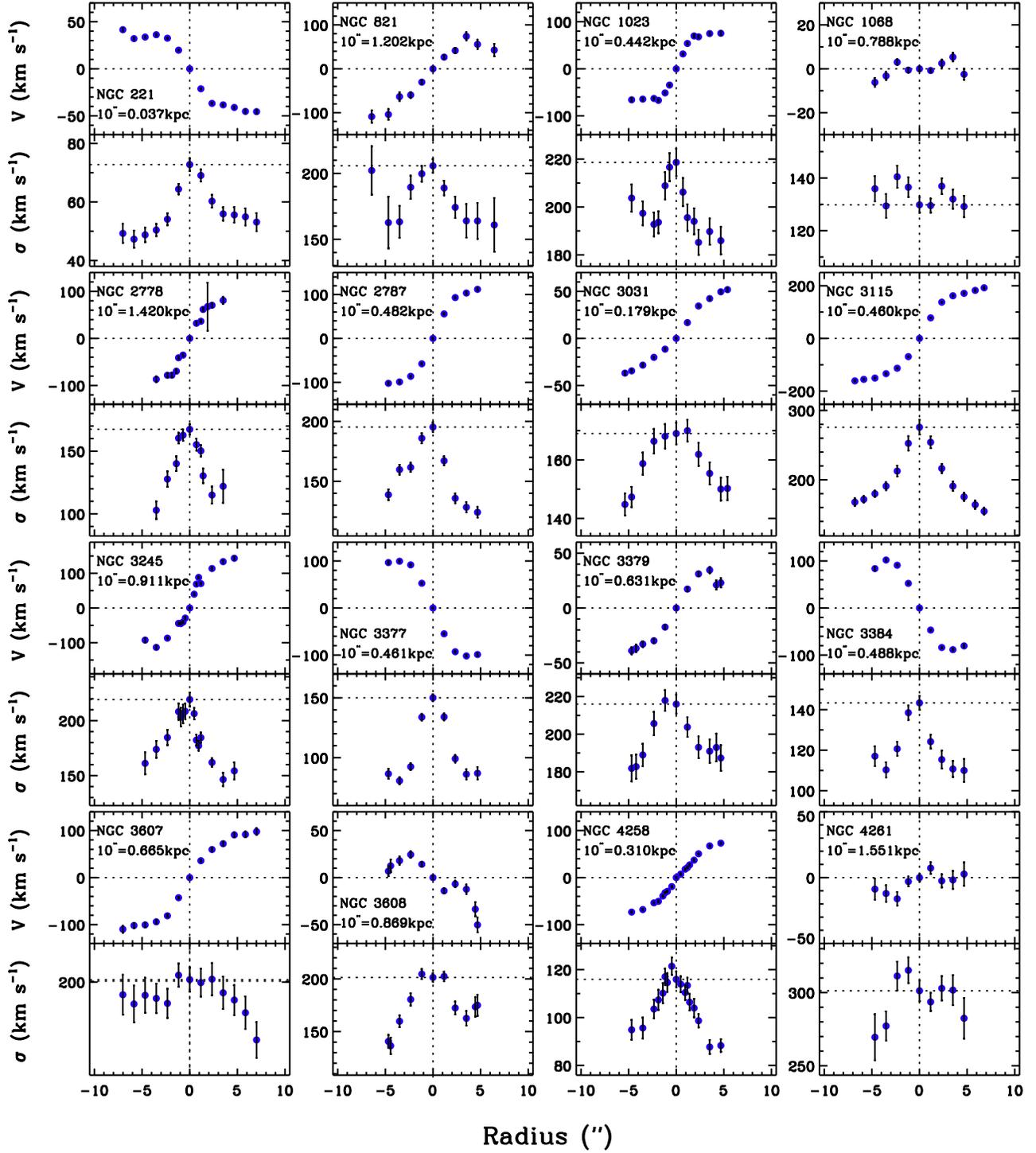}
\caption{The line-of-sight velocities (upper panel) and stellar velocity dispersions (lower panel) along the major axis. Most galaxies show a clear rotation component and a radial decrease 
of $\sigstar$. The object name and the spatial scale are shown in each upper panel.}
\label{fig4}
\end{figure*}

\begin{figure*}[htp]
\centering
\includegraphics[width = \textwidth]{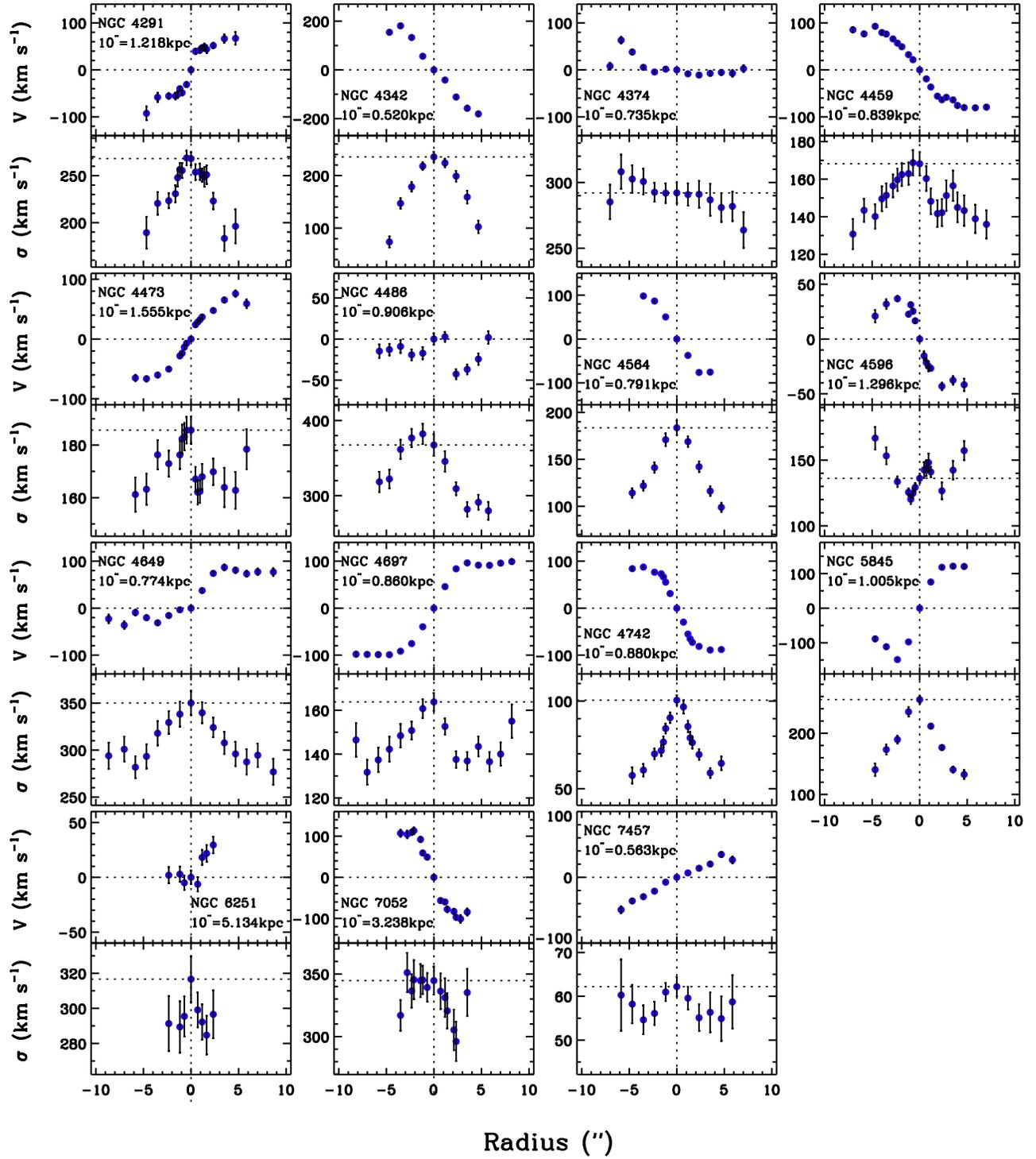}
\caption{Same as Figure \ref{fig4} for the rest of the sample.}
\label{fig5}
\end{figure*}

By extracting spectra over several bins along the major axis, we obtained spatially resolved kinematics. In Figures~\ref{fig4} and  \ref{fig5} we show the radial profiles of
line-of-sight velocities (upper panel) and velocity dispersions (lower panel) for each galaxy. 
We used the line-of-sight velocity of the galaxy center as a reference and normalized
all velocities with respect to the central value.
For most galaxies we extracted 9-13 spectra along the slit (in the direction of the major axis)
out to $\pm7\arcsec$ from the center. This is smaller than the slit size ($\pm15\arcsec$)
since we were not able to use outer pixels due to much lower S/N than the central bins to measure $\sigstar$
and the ABBA dither pattern along the slit.

Among the sample galaxies, we find a clear rotation component for 25 out of 31 objects. 
The amplitude of the projected rotation velocity ranges from $\sim$40 km s$^{-1}$ to over 
200 km s$^{-1}$ while six galaxies, namely NGC 1068, NGC 3608, NGC 4261, NGC 4374, NGC 4486 and 
NGC 6251 show a weak or no rotation component. 
For the galaxies with a significant rotation 
component, we expect line broadening due to the rotation, leading to overestimation of 
$\sigstar$, if a large single aperture is used for extraction. 
While the magnitude of this effect depends on the details of radial profiles of
rotation and velocity dispersion of each galaxy, it will lead to a systematic bias if 
not taken into account. We will account for the rotation component in the stellar velocity dispersion measurement in the next subsection.

%%% Figure 4%%%%%%%%%%%%%%%%%%%%%%%%%%%%%%%%%%%%%%%%%%%%%%%%%
%%%%%%%%%%%%%%%%%%%%%%%%%%%%%%%%%%%%%%%%%%%%%%%%%
%%% Figure 5%%%%%%%%%%%%%%%%%%%%%%%%%%%%%%%%%%%%%%%%%%%%%%%%%
%%%%%%%%%%%%%%%%%%%%%%%%%%%%%%%%%%%%%%%%%%%%%%%%%

%%% Figure 6%%%%%%%%%%%%%%%%%%%%%%%%%%%%%%%%%%%%%%%%%%%%%%%
\begin{figure*}[ht]
\centering
\includegraphics[width = \textwidth]{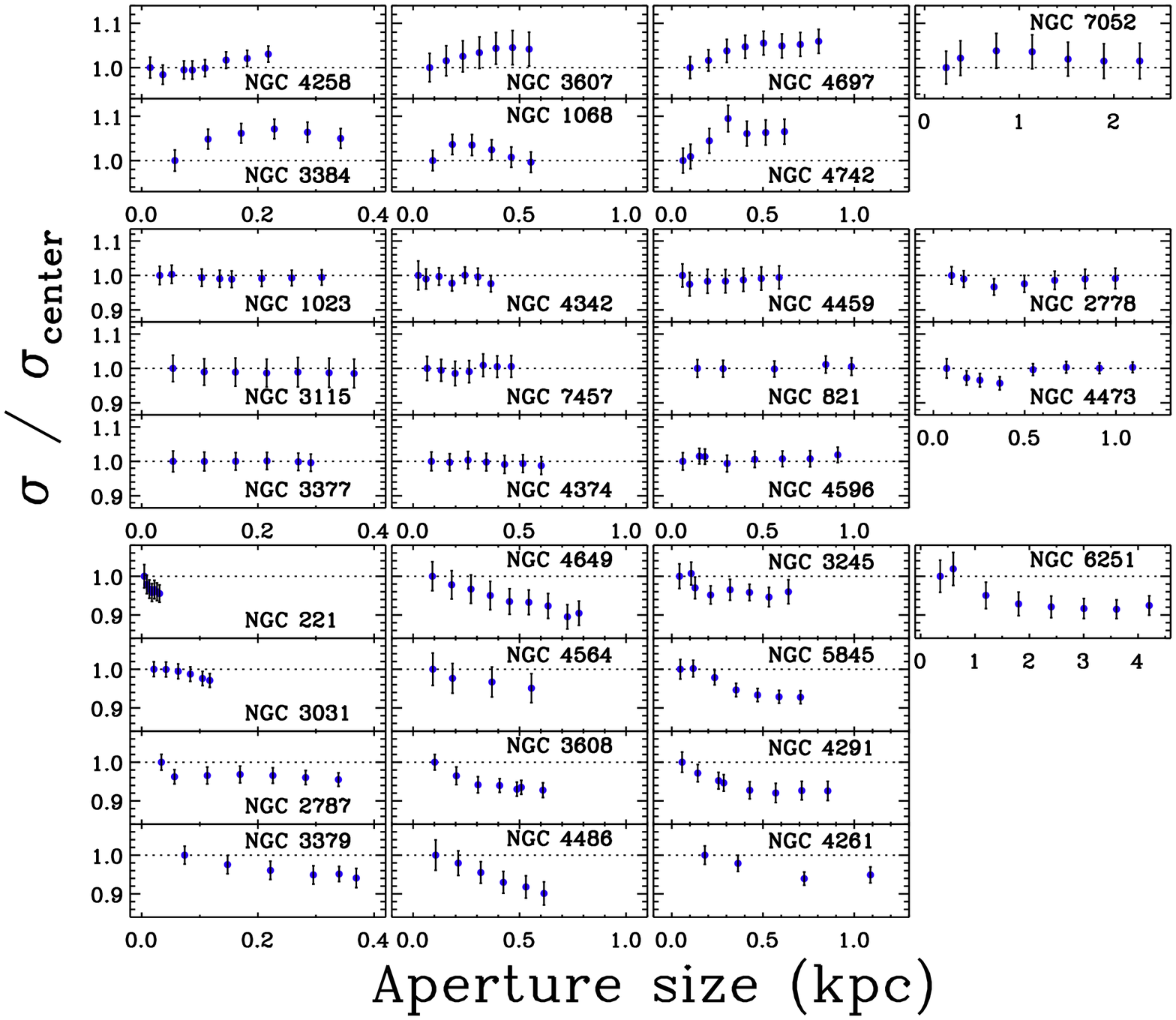}
\caption{Stellar velocity dispersions measured from various aperture sizes. Each $\sigstar$ measurement is normalized by $\sigstar$ measured from the smallest aperture. Upper panels show the objects that have increasing trend while Lower panels show the objects with an opposite trend.  Middle panels show the objects that have constant trend of $\sigstar$. Galaxies are sorted by the distance.}
\label{fig6}
\end{figure*}
%%%%%%%%%%%%%%%%%%%%%%%%%%%%%%%%%%%%%%%%%%%%%%%%%%%%%%

%We have classified the galaxies based on their stellar velocity dispersion profile:
%%%not touched!
%\textbf{NGC 221, NGC 1023, NGC 2778, NGC 2787, NGC 3379, NGC 3384, NGC 4258, NGC 4291 and NGC 4742} 
%- These galaxies show comparatively small deviation but the central parts have pointed shape. NGC 4291 shows asymmetric shape but judging from its error, the measurement of outer bin is uncertain.

In the lower panels of Figure \ref{fig4} and Figure \ref{fig5},  we show the stellar velocity 
dispersion profiles along the major axis. While stellar velocity dispersion decreases 
from the center to the outer regions for most galaxies,
several galaxies, e.g., NGC 1068, NGC 4261, NGC 4374, NGC 4596 and NGC 7052, 
do not show such a decreasing trend of $\sigstar$, 
but rather show irregular shapes; flat, increasing or asymmetric trends
as similarly reported by previous studies based on optical kinematics studies
\citep[e.g.][]{dress84, BSG94, kent90, pink+03}. 

\subsection{Aperture Size Effect}

%%% Figure 7%%%%%%%%%%%%%%%%%%%%%%%%%%%%%%%%%%%%%%%%%%%%%%%%%
\begin{figure}[ht]
\centering
\includegraphics[width = 0.47 \textwidth]{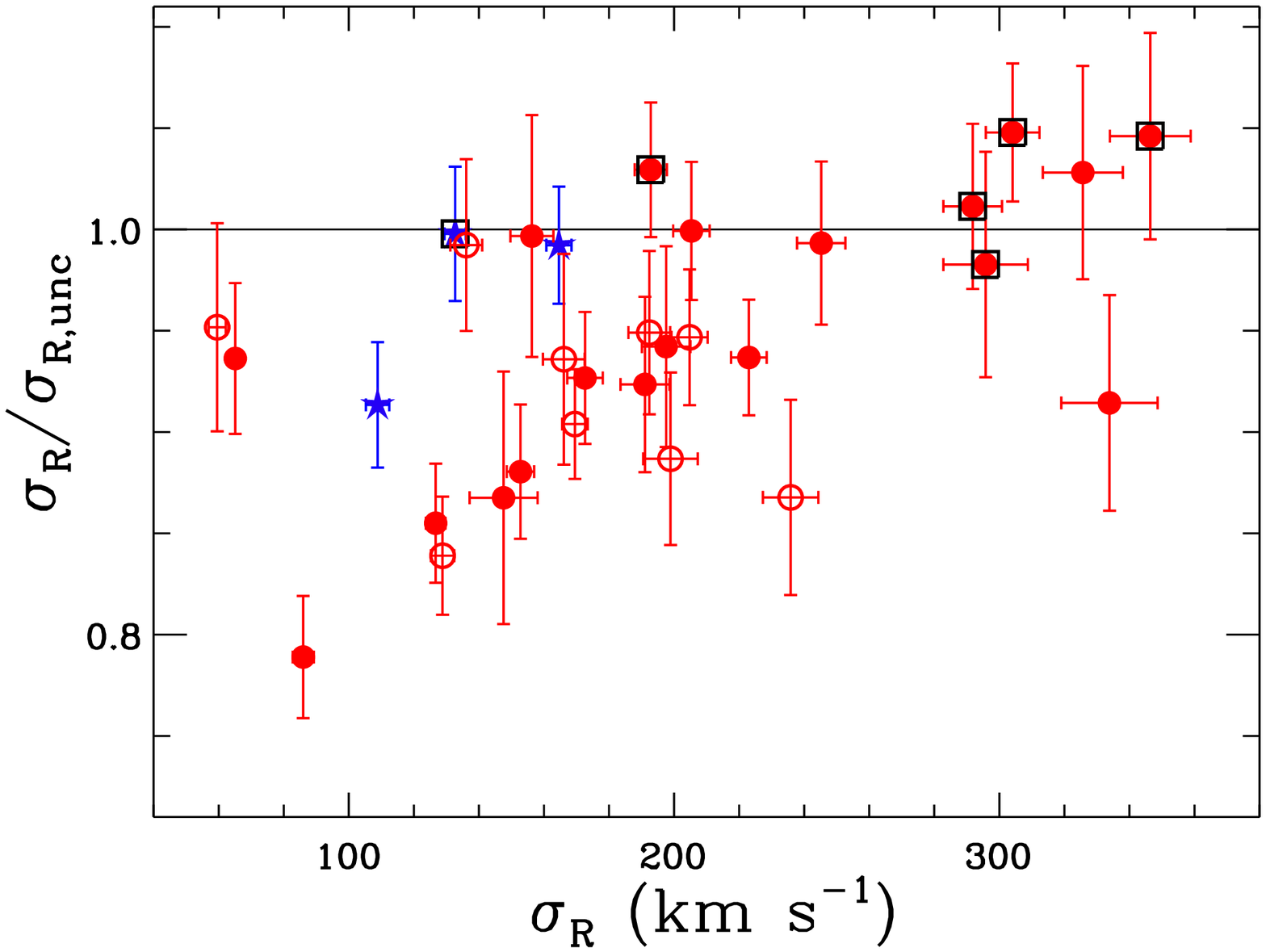}
\caption{Ratio of velocity dispersions determined with and without rotation correction as a 
function of velocity dispersion. Filled and open circles represent elliptical and lenticular galaxies,
respectively, while stars denote late-type galaxies. Six galaxies without a clear rotation component
are marked with open squares.}
\label{fig7}
\end{figure}
%%%%%%%%%%%%%%%%%%%%%%%%%%%%%%%%%%%%%%%%%%%%%%%%%%%%%%%%

We investigate the effect of using different aperture sizes on the measured stellar velocity dispersion, by directly measuring $\sigstar$ from apertures of increasing size. 
In Figure \ref{fig6} we present the $\sigstar$ measurements as a function of aperture
size, after normalizing them to the $\sigstar$ measured from the smallest aperture (4-10'').
We find three different trends (increasing, flat and decreasing) of $\sigstar$ with increasing aperture 
size. The 8 galaxies shown in the upper panels exhibit an increase of $\sigstar$
as a larger aperture size is used, while for 13 galaxies shown in the lower panels, $\sigstar$ decreases with increasing aperture size. These galaxies show variation of $\sigstar$ up to 20\% as aperture size changes. In contrast, 10 galaxies (middle panels)
do not show clear change of $\sigstar$ as a function of aperture size.

Thus, when measuring $\sigstar$ from a large aperture it is possible to either 
overestimate or underestimate $\sigstar$. The magnitude and direction of this bias 
depends on two factors: (1) the overestimation caused 
by rotational line broadening and (2) the natural decrease of $\sigstar$ as a function 
of radius. The galaxies in the upper panels in Figure \ref{fig6} are dominated by the 
first effect. They show relatively strong galaxy rotation and only a mild decrease in 
their velocity dispersion profile, 
as shown in Figures \ref{fig4} and \ref{fig5}, 
leading to a net increase in $\sigstar$ with increasing aperture size. 
For these galaxies, when the extraction aperture covers outer parts of the galaxy, 
where the rotation curves flatten, the aperture effect on $\sigstar$ also flattens. 
This is clearly seen for example in the case of NGC 3384, 
which shows a flattening of $\sigstar$ beyond the third bin, corresponding to the 
flattening of the rotation curve in Figure \ref{fig4}. 
NGC 4742 even shows a decrease in $\sigstar$ beyond the radius where the 
rotation curves becomes flat.

The decreasing $\sigstar$ trend for the galaxies in the lower panels of Figure \ref{fig6}
can be explained in a similar way. For these galaxies the decrease in $\sigstar$ profile 
is dominating over line broadening due to rotation. For example, NGC 4486 shows the 
largest variation in $\sigstar$ as a function of aperture size 
since it has no significant rotation component 
while the velocity dispersion profile is strongly decreasing toward larger radii. 
Similarly, NGC 3608 and NGC 4649 also show strong decrease, which is dominated by the 
strong decrease in $\sigstar$ with radius. In contrast, there are galaxies, e.g., 
NGC~821, where the effect from the rotation curve and the decreasing $\sigstar$ profile 
are roughly of the same order, leading to a small net variation of $\sigstar$ measured 
from different apertures.

\subsection{Correction of the Galaxy Rotation Effect} \label{section:rocorr}

As discussed above, $\sigstar$ measured within a certain aperture will be susceptible to 
line broadening by galaxy rotation. In contrast, the spatially resolved stellar velocity 
dispersions, represented in Figure \ref{fig4} and Figure \ref{fig5}, do not
suffer rotational broadening.
Thus, we can use these measurements to compute a rotation-corrected  $\sigstar$. 
We compute luminosity-weighted $\sigstar$ within a radius $R$:
\begin{equation}
\sigma_{R}=\frac{\int_{-R}^{R} \sigstar(r)\, I(r)\,  {\rm d}r}{\int_{-R}^{R}  I(r)\,  {\rm d}r} \, \label{eq1}
\end{equation}
where $I(r)$ is the surface brightness profile of the galaxy and $R$ is the outer radius within which we compute $\sigstar$.  
Using the spectral images, we measure the surface brightness profile of each galaxy
by fitting with two Gaussian models, and use this fit to compute the luminosity- weight
for the $\sigstar$ measured at each radius.
We chose an outer radius R for each galaxy based on the $\reff$ (see Table 1).
For 20 galaxies, we were able to measure spatially resolved $\sigstar$ only
at the central parts, due to the limited spatial coverage and/or lower S/N at 
the outer part. Thus, we chose 1/8 of $\reff$ as an outer radius in Eq. 1.
For the other 11 galaxies, we measured $\sigstar$ over a larger fraction of $\reff$ (1/4 to unity) as an outer radius and corrected for the rotation component as listed in Table 2.
%For NGC 2787, NGC 4342 and NGC 5845, we used $\reff$, 
%for NGC 3245, NGC 3384, NGC 4473, and NGC 4596, we used $\reff$/2, 
%and for NGC 2778, NGC 4258, NGC 4291, NGC 4742, we used $\reff$/4 
% reinstead of $\reff$, 

%%% Table 2%%%%%%%%%%%%%%%%%%%%%%%%%%%%%%%%%%%%%%%%%%%%%%%%%%%%%%%%
\begin{deluxetable}{ccccccc}[ht]
%\rotate
\tablecolumns{7}
\tablewidth{0pc}
\tablecaption{Near-IR and optical stellar velocity dispersions}
\tablehead{
\colhead{Name} &\multicolumn{4}{c}{$\sigma_{\rm IR}$} & \colhead{$\sigma_{\rm opt}$} & \colhead{Ref.}\\
\cline{2-5}
\colhead{} &\colhead{$\sigma_{\rm R, unc}$} & \colhead{$\sigma_{\rm R}$} & \colhead{$R$} & \colhead{$\sigma_{\rm \pm 7\arcsec}$} &\colhead{} &\colhead{}\\
\colhead{ }&\colhead{(km s$^{-1}$)}& \colhead{(km s$^{-1}$)}& \colhead{($\reff$)}& \colhead{(km s$^{-1}$)}&\colhead{(km s$^{-1}$)}& \colhead{ }\\
\colhead{(1)}&\colhead{(2)}&\colhead{(3)}&\colhead{(4)}&\colhead{(5)}&\colhead{(6)}&\colhead{(7)}\\
}
\startdata
N221           &$70{\pm 2}$              & $65{\pm 2}$       & 1/8    &$70{\pm 2}$         & $75{\pm 3}$            &1, 2\\
N821           &$207{\pm 5}$            & $191{\pm 8}$    & 1/8    &$208{\pm 5}$       & $209{\pm 10}$       &3\\
N1023         &$216{\pm 5}$            & $205{\pm 6}$    & 1/4    &$217{\pm 5}$       & $205{\pm 10}$       &4\\
N1068         &$133{\pm 3}$            & $133{\pm 3}$    & 1/8    &$129{\pm 3}$       & $151{\pm 7}$         &5\\
N2778         &$170{\pm 4}$            & $148{\pm 11}$  & 1/4   &$161{\pm 4}$       & $175{\pm 8}$         &3\\
N2787         &$188{\pm 4}$            & $170{\pm 4}$    & 1       &$186{\pm 3}$       & $189{\pm 9}$         &6\\
N3031         &$166{\pm 3}$            & $165{\pm 4}$    & 1/8    &$157{\pm 3}$       & $143{\pm 7}$         &7\\
N3115         &$272{\pm 11}$          & $236{\pm 9}$    & 1/8   &$272{\pm 12}$     & $230{\pm 11}$       &7\\
N3245         &$203{\pm 5}$            & $192{\pm 6}$    & 1/2    &$206{\pm 7}$       & $205{\pm 10}$       &7\\
N3377         &$148{\pm 4}$            & $127{\pm 3}$    & 1/8    &$147{\pm 4}$       & $145{\pm 7}$         &8, 9\\
N3379         &$205{\pm 4}$            & $205{\pm 6}$    & 1/8    &$203{\pm 5}$       & $206{\pm 10}$       &10, 2\\
N3384         &$154{\pm 3}$            & $129{\pm 4}$    & 1/2    &$151{\pm 3}$       & $143{\pm 7}$         &3\\
N3607         &$210{\pm 8}$            & $198{\pm 7}$    & 1/8    &$210{\pm 8}$       & $229{\pm 11}$       &11\\
N3608         &$187{\pm 4}$            & $193{\pm 5}$    & 1/8    &$187{\pm 4}$       & $182{\pm 9}$         &3\\
N4258         &$119{\pm 3}$            & $109{\pm 4}$    & 1/4    &$111{\pm 2}$       & $115{\pm 10}$       &6\\
N4261         &$290{\pm 5}$            & $304{\pm 8}$    & 1/8    &$286{\pm 6}$       & $315{\pm 15}$       &12, 2\\
N4291         &$247{\pm 7}$            & $245{\pm 7}$    & 1/4    &$248{\pm 7}$       & $242{\pm 12}$       &3\\
N4342         &$224{\pm 5}$            & $199{\pm 8}$    & 1       &$224{\pm 5}$       & $225{\pm 11}$        &13, 2\\
N4374         &$289{\pm 7}$            & $292{\pm 7}$    & 1/8    &$290{\pm 8}$       & $296{\pm 14}$        &14, 2\\
N4459         &$157{\pm 7}$            & $156{\pm 7}$    & 1/8    &$164{\pm 6}$       & $167{\pm 8}$          &7\\
N4473         &$186{\pm 3}$            & $173{\pm 5}$    & 1/2    &$186{\pm 3}$       & $190{\pm 9}$          &3\\
N4486         &$331{\pm 11}$          & $346{\pm 12}$ & 1/8    &$331{\pm 11}$     & $375{\pm 18}$&15\\
N4564         &$177{\pm 7}$            & $166{\pm 6}$    & 1/8    &$175{\pm 7}$       & $162{\pm 8}$         &3\\
N4596         &$137{\pm 3}$            & $136{\pm 5}$    & 1/2    &$139{\pm 3}$       & $136{\pm 6}$         &7\\
N4649         &$317{\pm 11}$          & $326{\pm 13}$ & 1/8     &$327{\pm 11}$    & $385{\pm 19}$       &3, 15\\
N4697         &$174{\pm 4}$            & $153{\pm 4}$    & 1/8     &$172{\pm 4}$      & $177{\pm 8}$         &3\\
N4742         &$109{\pm 3}$            & $86{\pm 3}$      & 1/4     &$104{\pm 3}$      & $90{\pm 5}$            &7\\
N5845         &$238{\pm 4}$            & $223{\pm 5}$    & 1        &$237{\pm 4}$      & $234{\pm 11}$       &9\\
N6251         &$301{\pm 11}$          & $296{\pm 13}$ & 1/8     &$290{\pm 8}$      & $290{\pm 14}$       &17, 2\\
N7052         &$365{\pm 14}$          & $334{\pm 15}$ & 1/8     &$327{\pm 13}$    & $266{\pm 13}$       &18\\
N7457         &$63{\pm 2}$               & $60{\pm 3}$      & 1/8    &$63{\pm 2}$         & $67{\pm 3}$            &3\\
\enddata
\label{table2}
\tablecomments{Col. (1): Object name. Col. (2): $H$-band stellar velocity dispersion measured using
a single aperture size $R$ without rotation correction. Col. (3): luminosity-weighted $H$-band $\sigstar$ within R. Col. (4): Aperture radius $R$ used for $H$-band $\sigstar$ in units of $\reff$. Col. (5): $H$-band velocity dispersion measured using a single aperture within $
{\pm7}\arcsec$. Col. (6):  optical stellar velocity dispersions. Col. (7): Reference for optical velocity dispersions.\\
References. --- (1) \citet{van98}; (2) \citet{geb+00}; (3) \citet{pink+03}; (4) \citet{bower+01}; (5) \citet{NW95}; (6) \citet{gul+09a}; (7) \citet{KG+01}; (8) \citet{kor98}; (9) \citet{geb+03}; (10) \citet{geb+00b}; (11) \citet{gul+09b}; (12) \citet{VBD90}; (13) \citet{CvdB99}; (14) \citet{bower98}; (15) \citet{geb+11}; (16) \citet{McC+11}; (17) \citet{SHI90}; (18) \citet{VV95}
}
\end{deluxetable}

In Figure \ref{fig7} we illustrate the effect of the correction for galaxy rotation. Here, $\sigcorr$ is  the luminosity-weighted $\sigstar$ within $R$ as computed from Equation~\ref{eq1} while $\siguncorr$ is measured from a single aperture with an aperture size of 
$R$. As expected, most galaxies show a decrease in velocity dispersion 
when accounting for the rotation component while for galaxies without strong
rotation component, the correction is marginal. 
Including 6 galaxies that show no rotation, the average correction is 6\%,
while the correction for individual galaxies can be up to $\sim$20\%.
The magnitude of the rotation correction tends to be smaller for more massive galaxies. 
NGC 7052 with the highest $\sigstar$ in the sample seems to be an outlier from 
this trend since it has relatively large rotation while $\sigstar$ mildly decreases
within $\reff$/8.  
In summary, we find that stellar velocity dispersions measured from single-aperture 
spectra can be biased by up to $\sim 20\%$. This is consistent with the results of \citet{bennert+11a} and \citet{harris+12}.

A possible drawback for the comparison with previous studies, that usually report $\sigstar$ measured within $\reff$, is the limited spatial coverage in our work, restricting our measurements to $\reff$/8. To investigate the effect on the $\sigstar$  measurement, we tested two extreme cases. First, we assumed a constant $\sigstar$ from $\reff$/8 to $\reff$, equal to the value at $\reff$/8. Second, we extrapolated the decreasing stellar velocity dispersion profile out to $\reff$. For both cases we computed the luminosity-weighted $\sigstar$ within $\reff$ via Equation~\ref{eq1}. We found that $\sigstar$ values decrease by only a 
few per cent by increasing the outer radius from $\reff$/8 to $\reff$ in both
cases, due to the much lower luminosity weight at outer radii. 
Thus, our $\sigstar$ measurements within $\reff$/8 will closely 
resemble the value that would be measured at $\reff$.

%==============================================================================================================

%%% Figure 8%%%%%%%%%%%%%%%%%%%%%%%%%%%%%%%%%%%%%%%%%%%%
\begin{figure}[ht]
\centering
\includegraphics[width = 0.42 \textwidth]{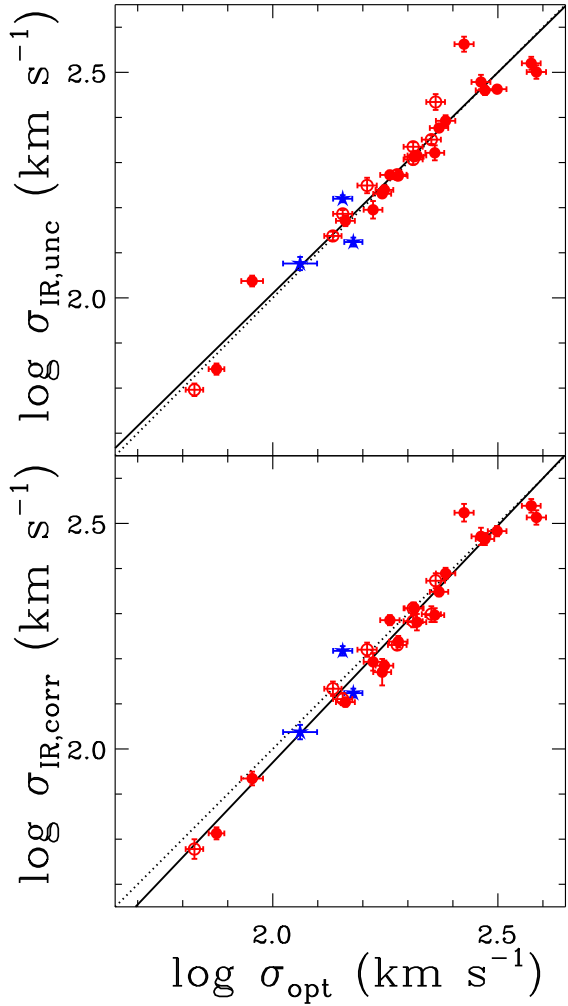}
\caption{Comparison between $H$-band stellar velocity dispersions and optical stellar 
velocity dispersions from the literature. $H$-band $\sigstar$ is measured from
an extraction aperture, ${\pm 7\arcsec}$ (top panel) or corrected for rotation effect 
(bottom panel). Elliptical and lenticular galaxies
are denoted with filled and open circles while 3 late-type galaxies are indicated by filled stars.
The best-fit (solid line) is consistent with a one-to-one relation (dotted line).}
\label{fig8}
\end{figure}
%%%%%%%%%%%%%%%%%%%%%%%%%%%%%%%%%%%%%%%%%%%%%%%%%%%%

\section{Discussion}
\subsection{Optical vs. near-IR Velocity Dispersions}

In Figure 8 we compare the stellar velocity dispersions measured using our near-IR 
spectra ($\sigir$) with the literature values measured from optical spectra ($\sigopt$).
For this comparison, we collected $\sigopt$ measurements from \citet{McC+13}, 
who listed their own measurements as well as previously measured 
values from the literature (see Table 2). Although these optical measurements were 
based on spatially resolved stellar kinematics and the quoted values were 
luminosity-weighted or averaged velocity dispersions within the $\reff$ for most galaxies,
these values were not homogeneously measured due to the various data quality and the 
measurement methods. In addition, some of these values in the 
original works were measured with a smaller aperture size than $\reff$ or the aperture
size was not clearly stated for many cases in the original references.

Thus, we decide to compare both $H$-band $\sigstar$ measurements with/without rotation correction
to the optical $\sigstar$ measurements. 
As shown in Figure \ref{fig8}, we find no significant difference between $\sigopt$ 
and $\sigir$. The best fit between optical and near-IR measurements
is close to a one-to-one relation with a scatter of $\sim$0.04 dex (10\%) when rotation
is not corrected for $\sigma_{IR}$.
The lower panel in Fig. 8 shows that the rotation-corrected $\sigma_{IR}$ is slightly
smaller than the optical $\sigstar$ at lower mass range. However, the average
offset is only 7\%, which is not significant compared to the measurement uncertainties
of stellar velocity dispersions. 

Note that \citet{McC+13} included rotation in calculating luminosity-weighted
$\sigstar$ by adding rotation velocity to velocity dispersion
in quadrature (See their Eq. 1). Thus, the slight offset between our $\sigma_{IR}$
and optical $\sigstar$ from \citet{McC+13} may be explained by the rotation effect.
To test this hypothesis, we derive rotation-included velocity dispersions using the same integral
as adopted by \citet[][Eq. 1]{McC+13}.
For these consistently measured velocity dispersions, we find that optical and IR velocity dispersions
show a one-to-one relationship with a slope of $1.00\pm0.05$ and a 0.03 dex (7\%) intrinsic scatter. 
Thus, we conclude that $\sigstar$ measurements derived from optical and H-band stellar lines are
consistent.

While many studies have been devoted to measure $\sigstar$ of galaxies using optical spectra,
the number of $\sigstar$ studies based on near-IR data, either $H$-band or $K$-band spectra is growing.
However, there are currently only few studies that actually compared the results from both wavelength 
regimes. For example, \citet{SG+03} measured $\sigstar$ of a sample of 25 elliptical and lenticular galaxies 
using the 2.29$\micron$ CO(2-0) band head in the $K$-band spectra. 
Comparing their IR results to optical velocity dispersions from the literature, they concluded 
that IR stellar velocity dispersions can be lower than optical stellar velocity dispersions, 
by up to 30$\%$ -- 40$\%$ and with a median offset of 11$\%$.
The inconsistency between optical and near-IR measurements is probably due to a sample bias and
measurement uncertainties. First, their sample mainly consists of S0 galaxies
and the systematic difference between optical and near-IR velocity dispersion in their study 
is mainly caused by S0 galaxies while their elliptical subsample does not show a difference between 
optical and near-IR measurements.
Second, \citet{SG+03} measured $\sigstar$ using solely a single CO band head in the $K$-band, 
which is much more susceptible to template mismatch as explained in their analysis.

In contrast to \citet{SG+03}, \citet{r&f10} reported no inconsistency between
optical and near-IR stellar velocity dispersion measurements for elliptical galaxies. 
Using a sample of 23 elliptical galaxies and 14 merger remnants, they measured $\sigstar$ 
from stellar lines in the $K$-band spectra, i.e., CO (2-0), CO (3-1), and CO (4-2) band heads, 
and compare them with velocity dispersion 
measured from the optical CaII triplet line, showing that optical and near-IR
stellar velocity dispersions are virtually the same for elliptical galaxies.
For merger remnants \citet{r&f10} reported a discrepancy between optical and near-IR velocity dispersions, 
presumably due to the presence of young stellar population, which are obscured at optical wavelengths.
However, for elliptical galaxies, their results indicate that optical and near-IR stellar lines 
represent the same kinematics and a dust effect is negligible.
Similarly, a recent study by \citet{vdb+11} presented near-IR $\sigstar$
measurements also based on the CO band heads for a sample of 22 galaxies, 
consisting of similar numbers of ellipticals and lenticulars. Comparing with previous optical 
measurements they reported that optical and near-IR $\sigstar$ were consistent for their sample,
which is consistent with our results. 

In the case of velocity dispersions using $H$-band stellar lines,
there has been no systematic comparison with optical velocity dispersions. 
Using various stellar lines in the $H$-band spectra and carefully accounting for the template mismatch 
problem (see Section 3.2), for the first time, we show that optical and $H$-band $\sigstar$ measurements 
are consistent for early-type galaxies, indicating that optical and $H$-band stellar lines represent
the same kinematics and that a dust effect, i.e., obscuration at optical wavelengths,
is negligible. These results are consistent with $K$-band stellar kinematics \citep{r&f10, vdb+11}.
Our results imply that near-IR $\sigstar$ measurements carried out for AGN host galaxies, for which 
optical measurements are more difficult to perform due to the strong AGN contribution
(e.g., Woo et al. 2010), provide unbiased results, compared to optical measurements.

%==============================================================================================================

%%% Figure 9%%%%%%%%%%%%%%%%%%%%%%%%%%%%%%%%%%%%%%%%%%%%%%%%%%%%%%
\begin{figure}
\centering
\includegraphics[width = 0.47 \textwidth]{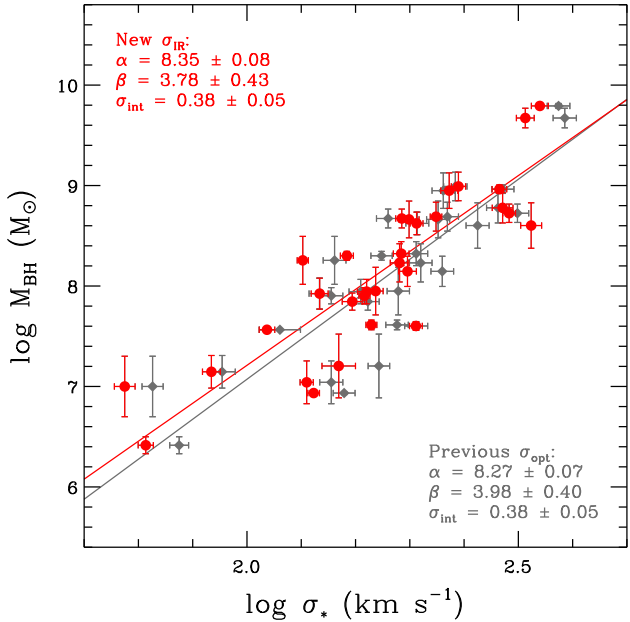}
\caption{The \msigma\ relation of 31 nearby galaxies, using rotation-corrected $\sigstar$ measured from
our H-band spectra (red circles, red line) and rotation-included optical $\sigstar$ from
\citet{McC+13} (gray diamonds, gray line), respectively.}
\label{fig9}
\end{figure}
%%%%%%%%%%%%%%%%%%%%%%%%%%%%%%%%%%%%%%%%%%%%%%%%%%%%%%

\subsection{The \msigma\ Relation for Early-type Galaxies} \label{section:msig}

In this paper we present homogeneously measured $\sigstar$ for 31 galaxies with dynamical $\mbh$ measurements. By accounting for galaxy rotation and implementing a uniform analysis for measuring velocity dispersions, 
our $\sigstar$ measurements are slightly different from previous optical values. 
In this section we demonstrate the effect of these new velocity dispersions on the \msigma\ relation 
by fitting the \msigma\ relation for 31 galaxies, for which we obtained the rotation-corrected $\sigstar$. 
Results on the  \msigma\ relation for the full sample of galaxies with dynamical $\mbh$ will be 
presented in a companion paper (Woo et al. 2013 in preparation).
For this analysis, we used the most recent $\mbh$ updates from \citet{McC+13}.

We fit the \msigma\       relation as a single-index power law:
\begin{equation}
 \log (\mbh/ M_\odot ) =\alpha + \beta \log (\sigstar/ 200\,\mathrm{km\,s}^{-1}) \ .
\end{equation}  
We used the FITEXY method, modified to account for intrinsic scatter in the relation 
\citep{T+02,park+12}, to perform the fit as shown in Figure \ref{fig9}.

By fitting the \msigma\ relation using the optical $\sigstar$ from \citet{McC+13},
we obtain  $\alpha=8.27 {\pm 0.07}$, $\beta=3.98 {\pm 0.40}$ and an intrinsic scatter of $0.38 {\pm 0.05}$ dex. 
Our sample has a large overlap with the sample used by \citet{T+02}, and indeed we obtain consistent results for the \msigma\ relation. 
The slight difference arises mainly from a few different galaxies in 
both samples and from updated $\mbh$ determinations \citep[e.g.][]{SG+11}.
By fitting the \msigma\       relation using our rotation-corrected near-IR $\sigstar$,
we find  $\alpha=8.35 {\pm 0.08}$, $\beta=3.78 {\pm 0.43}$ and an intrinsic scatter of $0.38 {\pm 0.05}$ dex.
This relation is slightly shallower than that derived from optical $\sigstar$, but 
consistent within the uncertainties. 

Note that the previous studies on the \msigma\
relation by \citet{gul+09a} and \citet{McC+13} explicitly included rotation in
calculating luminosity-weighted $\sigstar$.
In this case, we expect systematic effect on the measured $\sigstar$ due to the random
orientation of stellar disk with respect to the line-of-sight. To investigate this effect,
we calculated $\sigstar$ by adding velocity to velocity dispersion in quadrature 
using Eq. 1 in \citet{McC+13}.
The result shows that rotation-included $\sigstar$ is slightly larger than rotation-corrected 
$\sigstar$, particularly at low mass scale, by 0.02 dex ($\sim$5\%) on average with a 
0.02 ($\sim$5\%) scatter. Consequently, when we replace rotation-corrected $\sigstar$
with rotation-included $\sigstar$ in fitting the the \msigma\ relation, the slope slightly 
increases from $3.78\pm0.43$ to $3.97\pm0.49$ as the $\sigstar$ values increase preferentially
at low mass scale, while intrinsic scatter remains the same.

Although we expect that the rotation effect will systematically affect the \msigma\ relation,
we do not clearly detect the improvement of the \msigma\ relation by using rotation-corrected $\sigstar$,
presumably due to two reasons. 
First, rotation effect on the luminosity-weighted velocity dispersion may not be significant 
as the integrated velocity dispersions are dominated by the inner part, where 
the rotation velocity is relatively small. In the case of the rotation-included velocity dispersion, 
the luminosity weight of the inner part is more dominant since velocity dispersions are integrated 
in quadrature \citep[see Eq. 1 in][]{McC+13}. 
Secondly, since our sample is mainly composed of early-type galaxies,
rotation effect is relatively weak compared to late-type galaxies. For 
late-type galaxies with a low $\sigstar$, rotation effect can be significant, hence, 
it would be essential to correct for, in order to properly derive the \msigma\ relation. 

For massive BHs, the sphere of influence of BH can be large enough to change the 
effective $\sigstar$ measurements since the velocity dispersion at the center
increases due to the presence of a BH. Thus, by excluding the sphere of influence of BH
in calculating the luminosity-weighted $\sigstar$, the effective velocity dispersions will be decreased. 
Since these corrections can be done only for massive galaxies with a resolved sphere of influence,  
the slope of the \msigma\ relation will increase due to the preferential decrease of $\sigstar$ at high
mass end. For example, by excluding the 
sphere of influence of BH in deriving the effective $\sigstar$ within R$_e$ for 12 most massive 
galaxies, \citet{McC+13} showed that the slope of the \msigma\ relation increased from $5.48\pm0.30$ to $5.64\pm0.32$. 
We performed a similar analysis using our data although only two galaxies, NGC 4486
and NGC 4649, are among those 12 galaxies with a resolved sphere of influence. 
By excluding the sphere of influence of BH,
the luminosity-weighted $\sigstar$ decreases from $346\pm12$ to $327\pm11$ for NGC 4486, and from
$346\pm12$ to $327\pm11$ for NGC 4649. Based on these two updated $\sigstar$, the slope
of the \msigma\ relation slightly increases from $3.78\pm0.43$ to $3.79\pm0.45$, however
two slopes are consistent within the uncertainties.
Using only 2 galaxies, it is not clear whether excluding or including the sphere of
influence in determining the effective $\sigstar$ improves the \msigma\ relation.

Compared to the  \msigma\   relation recently presented by \citet{McC+13}, we find 
a significantly shallower slope. They report a slope of $5.64\pm0.32$, using a much larger 
galaxy sample, which includes in particular more galaxies at higher and lower masses. 
We will investigate the implications of our results on the \msigma\         relation in detail 
in a companion paper (Woo et al. 2013 in preparation).

%==============================================================================================================

\section{SUMMARY} \label{section:sum}

We observed a sample of 31 nearby galaxies with Triplespec, a near-IR long-slit spectrograph at the 
Palomar 5-m telescope in order to homogeneously measure velocity dispersions from the $H$-band stellar lines.
The galaxies in the sample cover a wide range in $\sigstar$ (67 km s$^{-1}$ $<$ $\sigstar$ $<$ 385 km s$^{-1}$) 
and their dynamical central BH masses are also available.
To account for template mismatch, we used 11 giant stars with spectral type ranging from K0 to M5
as velocity templates, and found that M giants generate the most reliable fits and velocity dispersion measurements.

By measuring velocity and velocity dispersion as a function of radius along the major axis of each galaxy,
we determined the rotation curve and velocity dispersion profile. 
Using these spatially resolved velocity dispersion measurements, 
we calculated the luminosity-weighted stellar velocity dispersions within the $\reff$ of each galaxy.
For 25 out of 31 galaxies in the sample, we found a clear rotation component, indicating that stellar 
velocity dispersions can be significantly overestimated due to the rotational broadening
if a large single aperture is used to extract spectra.
Compared to rotation-corrected velocity dispersions,
velocity dispersions measured from single-aperture spectra showed systematically larger values by up to $\sim$20$\%$.

We compared velocity dispersions measured from $H$-band stellar lines with those measured from optical lines
and found no systematic difference, suggesting that optical and $H$-band stellar lines represent
the same kinematics and that dust effect is negligible for early-type galaxies.
Our results confirm that optical and near-IR stellar lines can be interchangeably used to measure stellar kinematics
and near-IR $\sigstar$ measured for AGN host galaxies can be directly compared to optical $\sigstar$
of quiescent galaxies.

Using the rotation-corrected $\sigstar$ measurements based on the spatially-resolved H-band spectra
of 31 nearby galaxies, we derived the \msigma\ relation to investigate rotation effect.
The slope of the \msigma\ relation is slightly 
shallower than that based on the rotation-included optical or near-IR $\sigstar$ measurements.
Although rotation effect is not dramatically strong for early-type galaxies, it is potentially
important to correct for, particularly for low mass, late-type galaxies with a strong 
rotation component, in order to properly determine the \msigma\ relation and its intrinsic 
scatter. A future study based on spatially resolved spectra for late-type galaxies is
required to fully quantify rotation effect on the \msigma\ relation.

\acknowledgements
We thank the anonymous referee for constructive suggestions, which improved the manuscript.
This work was supported by the National Research Foundation of Korea (NRF) grant funded
by the Korea government (MEST) (No. 2012-006087).
J.H.W acknowledges the support by the Korea Astronomy and Space Science Institute (KASI) grant funded by the Korea government (MEST).

\clearpage

\end{document}